\documentclass[sigconf]{acmart}

\AtBeginDocument{%
  \providecommand\BibTeX{{%
    \normalfont B\kern-0.5em{\scshape i\kern-0.25em b}\kern-0.8em\TeX}}}
\acmBooktitle{Companion Proceedings of the 33rd ACM Symposium on the Foundations of Software Engineering (FSE '25), June 23--27, 2025, Trondheim, Norway}
\acmYear{2025}\copyrightyear{2025}
\setcopyright{rightsretained}
\acmConference[FSE '25]{33rd ACM International Conference on the Foundations of Software Engineering}{June 23--28, 2025}{Trondheim, Norway}
\acmBooktitle{33rd ACM International Conference on the Foundations of Software Engineering (FSE '25), June 23--28, 2025, Trondheim, Norway}
\acmDOI{10.1145/3696630.3731623}
\acmISBN{979-8-4007-1276-0/25/06}
\usepackage{tikz}
\usepackage{url}
\usetikzlibrary{chains,shapes.multipart}
\usetikzlibrary{shapes}
\usetikzlibrary{automata,positioning}
\usetikzlibrary{calc}
\usepackage{pgfplotstable}
\usepackage{pgfplots}
\usepackage{tikzscale}
\usepackage{makecell}
\usepackage{tabularx}
\usepackage{tkz-graph}
\usetikzlibrary{shapes.geometric}
\usepackage{listings}
\usepackage{./quantikz}
\usepackage{pifont}
\usepackage{caption}
\usepackage{subcaption}
\usepackage{easy-todo}
\usepackage{multirow}
\usepackage{colortbl}
\usepackage{color,soul}
\usepackage[most]{tcolorbox}
\usepackage{amsfonts}
\usepackage{multicol}
\usepackage{balance}
\usepackage{float}

\newcolumntype{C}[1]{>{\centering\let\newline\\\arraybackslash\hspace{0pt}}m{#1}}

\begin{document}

\title{The Road to Hybrid Quantum Programs: Characterizing the Evolution from Classical to Hybrid Quantum Software}


\author{Vincenzo De Maio}
\email{vincenzo.maio@tuwien.ac.at}
\affiliation{
  \institution{TU Wien}
  \city{Vienna}
  \country{Austria}
}
\author{Ivona Brandic}
\affiliation{
  \institution{TU Wien}
  \city{Vienna}
  \country{Austria}
}
\author{Ewa Deelman}
\affiliation{
  \institution{University of Southern California}
  \city{Los Angeles}
  \state{California}
  \country{USA}
}
\author{Jürgen Cito}
\affiliation{
  \institution{TU Wien}
  \city{Vienna}
  \country{Austria}
}




\begin{abstract}
Quantum computing exhibits the unique capability to natively and efficiently encode various natural phenomena, promising theoretical speedups of several orders of magnitude. However, not all computational tasks can be efficiently executed on quantum machines, giving rise to hybrid systems, where some portions of an application run on classical machines, while others utilize quantum resources. 
However, existing methods often to identify candidates to quantum execution (quantum candidates) involve a trial-and-error approach, relying on the intuition and expertise of computer scientists, resulting in varying identification durations ranging from minutes to days for a single application. 

This paper aims to systematically formalize the process of identifying quantum candidates and their proper encoding within classical programs. Our work addresses the critical initial step in the development of automated reasoning techniques for code-to-code translation, laying the foundation for more efficient quantum software engineering. 

\end{abstract}

\maketitle
\section{Introduction}

As we entered post-Moore Era, the computer science community is faced with new challenges~\cite{10.1145/216585.216588}: data volumes are growing faster than the computing power, forcing the community to look for alternatives beyond von Neumann architecture. Among different architectures that are currently being developed, Quantum Computing is one of the most promising. 
Quantum Computing (QC) aims to tackle a wide array of complex problems by leveraging its unique and novel features, such as superposition and entanglement, that enables solving problems that are not possible with classical computers.
These range from drug discovery~\cite{drugquantum}, manufacturing~\cite{manifacturingquantum}, healthcare~\cite{healthcarequantum}, finance~\cite{orus2019quantum} and many more ~\cite{bauer2020quantum, bharadwaj2020quantum, ollitrault2021molecular}. 
However, not all computational steps of an application can be executed efficiently on a quantum machine, requiring so-called hybrid systems where certain parts of the application are executed on the classic machine and other parts are executed on the quantum machine~\cite{Cranganore2022, weigold2021encoding}. 

The development of QC applications is a complex endeavor that requires experts of different fields to work in tandem, including domain experts such as chemists to properly understand the problem space and goals of the computation, computer scientists for expressing problems as efficiently executable programs, and physicists for mapping amenable sub-problems to quantum algorithms. Developing a hybrid application is an even more complex endeavor. Methods used to understand and develop hybrid systems range from automatic and semi-automatic code analysis to fundamental software engineering approaches. 
Static code analysis can be used to identify code fragments that represent quantum candidates. However, most static code analysis approaches are in its infancy~\cite{kaul2023uniform} with limited applicability (i.e., specific frameworks/languages~\cite{jimenez2020reverse}) and generalizability~\cite{ali2020modeling}. 
Beyond some early and preliminary work on static code analysis, the most common method to identify quantum candidates is a trial-and-error approach where domain experts and computer scientists identify quantum candidates in an exploratory fashion manually. 
The effort to identify quantum candidates depends on the experience of the computer/domain scientists and can last from minutes to days for the same application~\cite{DBLP:conf/supercomputer/MaioB23}.



We conducted a comprehensive six-month auto-ethnographic case study, engaging two distinct case partners – one, a seasoned domain expert in molecular dynamics, and the other, a team of quantum physicists and quantum computing scientists. The primary objective of this study was to document and formalize the intricate process of transitioning a conventional scientific program, rooted in the realm of molecular dynamics, into a hybrid quantum program, leveraging IBM's quantum machines through the Qiskit framework. Our approach of triangulating data sources, comprising detailed personal notes by the first author, version-controlled software evolution artifacts, and computational notebooks that emerged from collaborative efforts among project stakeholders, allowed us to gain invaluable insights into the software evolution journey. The findings we present in this paper outline the multifaceted steps required for the transition of classical programs into hybrid quantum programs. These steps encompass various phases, ranging from initial conceptual decomposition into tasks and identification of quantum candidates, through to their mapping into quantum subroutines, establishing of proper abstractions to ensure concurrent operation of classical and quantum variants, and culminating in the vital stages of integration and evaluation, facilitated by automated differential testing. Our study documents a comprehensive roadmap for the evolution of classic software into the quantum realm and formulates implications for quantum software engineering. Furthermore, we conduct a comparative analysis of our findings to discern how existing approaches align with our proposed quantum software evolution model.
We make the following contributions: (1) we characterize the process by which domain experts, in tandem with computer scientists, turn a classical application into a \emph{functionally equivalent} quantum hybrid program based on a case study on a migration of a molecular dynamics application; (2) for each process step, we identify a more abstract methodology and discuss how it can be generalized beyond our case. We also present the use case specifics of our case study alongside the methodology to anchor our conclusions, and (3) we map our findings on the process of quantum software evolution to existing and ongoing quantum software engineering research and sketch potential future avenues of investigation.

Different frameworks to enable the execution of computation on hybrid quantum systems are available, ranging from IBM Qiskit, Amazon BraKet, Google Cirq, Microsoft Azure Quantum, and PennyLane AI. We focus on IBM Qiskit, which is currently the most used framework in this context and allows users from different domains to run applications on hybrid classic/quantum systems using IBM classic and quantum hardware. 



\section{Related Work}


\textbf{Quantum Software Engineering.} 
There is a plethora of recent papers that investigate the use of software engineering techniques to quantum programs~\cite{luo:22, destefano:22, weder:22, zhao:21, paltenghi:22, ali:21, wang:21, fortunato:22, ahmad:22, li:21, akbar:22, garcia:21, quanfuzz}. The most prominent research area for quantum software engineering is software testing of quantum programs~\cite{ali:21, fortunato:22, quanfuzz, wang:21}. Ali et al. investigate the efficacy of input and output coverage criteria when testing quantum programs~\cite{ali:21}. 
Fortunato et al. investigate how mutation testing can be applied for quantum programs~\cite{fortunato:22}. They build on syntactically equivalent quantum operations to generate mutants based on qubit measurements and quantum gates. 
Wang et al. introduce QuanFuzz, a tool designed for generating test inputs for quantum software using a search-based approach~\cite{quanfuzz}.
Qdiff uses differential testing to identify bugs between different quantum platforms~\cite{wang:21}. 
As part of formulating implications of our study, we mapped how current open challenges in the quantum software engineering evolution process map to these existing efforts (see Section \ref{sec:implications}).


\noindent\textbf{Quantum Applications.} Other existing work investigates how the use of quantum hardware can enhance the performance of existing classic applications. In~\cite{Chang2021}, authors define methods to implement the Independent Set problem on IBM Quantum machines. One of the most promising applications of quantum computing is the optimization of linear algebra problems~\cite{xu2021variational}. These problems involve large matrices and vectors computation and find their applications in a variety of fields such as machine learning~\cite{biamonte2017quantum}, cryptography~\cite{easttom2022quantum}, finance~\cite{orus2019quantum} and molecular dynamics~\cite{niklasson2012fast}.

Concerning other scientific computing use cases, In~\cite{Cranganore2022}, authors enhance molecular dynamics application by offloading part of computation on quantum hardware. In~\cite{Matic2022}, authors apply quantum machine learning to classify medical images, while~\cite{daley2022practical} focuses on quantum simulations. Integration of quantum elements in machine learning applications is discussed in~\cite{cerezo2022variational}. 

\section{Case Study}


This study aims to investigate how domain experts (i.e., scientists) and software developers take an existing scientific program written as an imperative program and turn it into an executable hybrid quantum program, mixing elements of classic and quantum computation by decomposing and transforming the initial program. This process is inherently social, involving experts in the application domain, quantum computing experts, and computer scientists.




We employ methods from ethnography to study to answer research questions.
Particularly, we perform an autoethnography~\cite{ellis:11}, a variation of traditional ethnography, which offers a distinctive research perspective by placing the researcher as both the subject and the analyst. In autoethnography, individuals explore their own experiences within a specific cultural or social context, providing a subjective but valuable viewpoint. This methodology is compelling for our case study because it allows researchers to offer firsthand, reflexive accounts of their experiences and perceptions.
More specifically, our research is guided by the principles of interpretive field research as part of a case study. 
Interpretive field research aims to understand the deeper structure of a phenomenon from a participant’s perspective~\cite{orlikowski1991studying}.
Our findings are derived through analysis of the source code of the existing scientific programs, the social process of identifying subproblems that are amenable to execution on quantum hardware ("quantum candidates"), and the software evolution that eventually leads to a hybrid quantum program. 
In that sense, we follow an interpretivist perspective as our knowledge is derived through social constructions ``such as language, consciousness, shared meanings, documents, tools, and other artifacts"~\cite{gasson2004rigor}.


\subsection{Research Site: A Molecular Dynamics Case}
We worked together with two partners, A and B, to perform our case study in the scientific computing domain.
Team A focuses on scientific workflows, particularly in molecular dynamics (MD) applications, which are amenable to quantum execution~\cite{Chang2021}. Team B are quantum physicists who work in quantum mechanics, specifically with applications to super-computing.

We focus on an application described in~\cite{Cranganore2022}. Team A developed the initial ``classical" scientific program. This program together with the experiences build the basis for our initial investigation.
Team B provided quantum computing expertise to support the evolution steps.
The first author is a computer scientist working in high-performance computing (HPC) and was an active member of the team as part of this case study. We make the resulting  application available as an artifact\footnote{This is an anonymized version of our artifact. \url{https://anonymous.4open.science/r/quantumMD-A70B/}}.

\subsection{Data Collection}

For this case study spanning six months, data collection was conducted using a multi-faceted approach to ensure the richness and depth of our findings. We initiated the study by collaborating with members of Team A, who possess extensive expertise in scientific applications, to identify a suitable target use case for our methodology. This initial consultation provided valuable insights into the practicality and relevance of our research.

To maintain an ongoing dialogue and facilitate the study's progress, weekly plenary meetings were held with either Team A or Team B, depending on whether we identified challenges with either the case application or quantum computing. These meetings served as a platform for tracking developments in the case study, addressing any emerging challenges, and fostering a collaborative environment for knowledge exchange.

Our data collection strategy employed several sources to triangulate findings and mitigate potential limitations inherent in the participant observation methodology.

\noindent\textbf{Personal Notes.} The first author, who actively participated in the research site, meticulously documented personal notes. These notes encompassed a comprehensive record of the research process, including interactions and communications with various stakeholders involved in the case. These personal notes provided an insider's perspective and helped to contextualize our findings.

\noindent\textbf{Documented Software Evolution.} To capture the evolution of the classical program into a hybrid quantum program, we relied on the version control history of the project. This history not only included the Python scripts representing the eventual hybrid quantum program (and intermediate states therein) but also computational notebooks (i.e., Jupyter notebooks), which combined narrative explanations with code explorations. Additionally, we examined commit messages associated with code changes to extract the rationale behind each modification.

\noindent\textbf{Computational Notebooks.} In addition to personal notes as a form of an ongoing record of the case, we leveraged the extensive narrativistic components of our computational notebooks that were the result of specialized investigations within the project (especially with Team B). These notebooks served as formalized documentation, facilitating the communication of complex concepts among stakeholders with varying areas of expertise. 

This data collection approach allowed us to capture the nuances of our case, ensuring a comprehensive exploration of the research context and the dynamic processes unfolding within it.

\subsection{Analysis}

Our analysis was split into two phases. In our first phase, we conduct \emph{thematic analysis}~\cite{thematic} on our collected data to identify overarching themes, resulting in the synthesis of findings (Section \ref{sec:findings}). 
We focused on iterative coding where we developed initial themes while going through the data, which we further split down and transformed along the
way. Repeating this process loop of open, axial, and selective coding several times allowed us to reduce the amount of data to reason about while putting the focus on the relevant parts and abstractions for our research questions.

This is followed by a theoretical analysis of the findings, stemming from a comprehensive thematic examination of ethnographic data, which seeks to establish a bridge between the observed outcomes and the pre-existing body of knowledge within the realm of quantum software engineering. This endeavor involves delving into the core themes and patterns extracted from our study and subsequently contextualizing them within the broader landscape of quantum software engineering literature. By mapping the identified nuances and insights onto the existing theoretical frameworks and paradigms, this analysis aims to highlight how the ethnographic findings contribute to and align with current understandings of quantum software engineering practices. In doing so, it not only enriches the scholarly discourse but also fosters a deeper comprehension of the intricate interplay between empirical observations and established theoretical underpinnings in this field.

\section{Findings}
\label{sec:findings}



After careful thematic analysis, we identified the following crucial steps involved in evolving classical programs into hybrid quantum programs: (\textit{i}) decomposition of application in sub-problems, (\textit{ii}) identification of quantum candidates, (\textit{iii}) data encoding, (\textit{iv}) implementation of quantum code, and (\textit{v}) integration in the hybrid quantum-classical loop. First, we provide an overview of the software development process, then we describe each step in the following sections. For each step, we carefully separate a more abstract, general methodology that we expect to generalize beyond our case to the use case specifics that were performed as part of the case study. 

\subsection{Overview}
\begin{figure}[H]
    \centering
\includegraphics[width=.98\columnwidth]{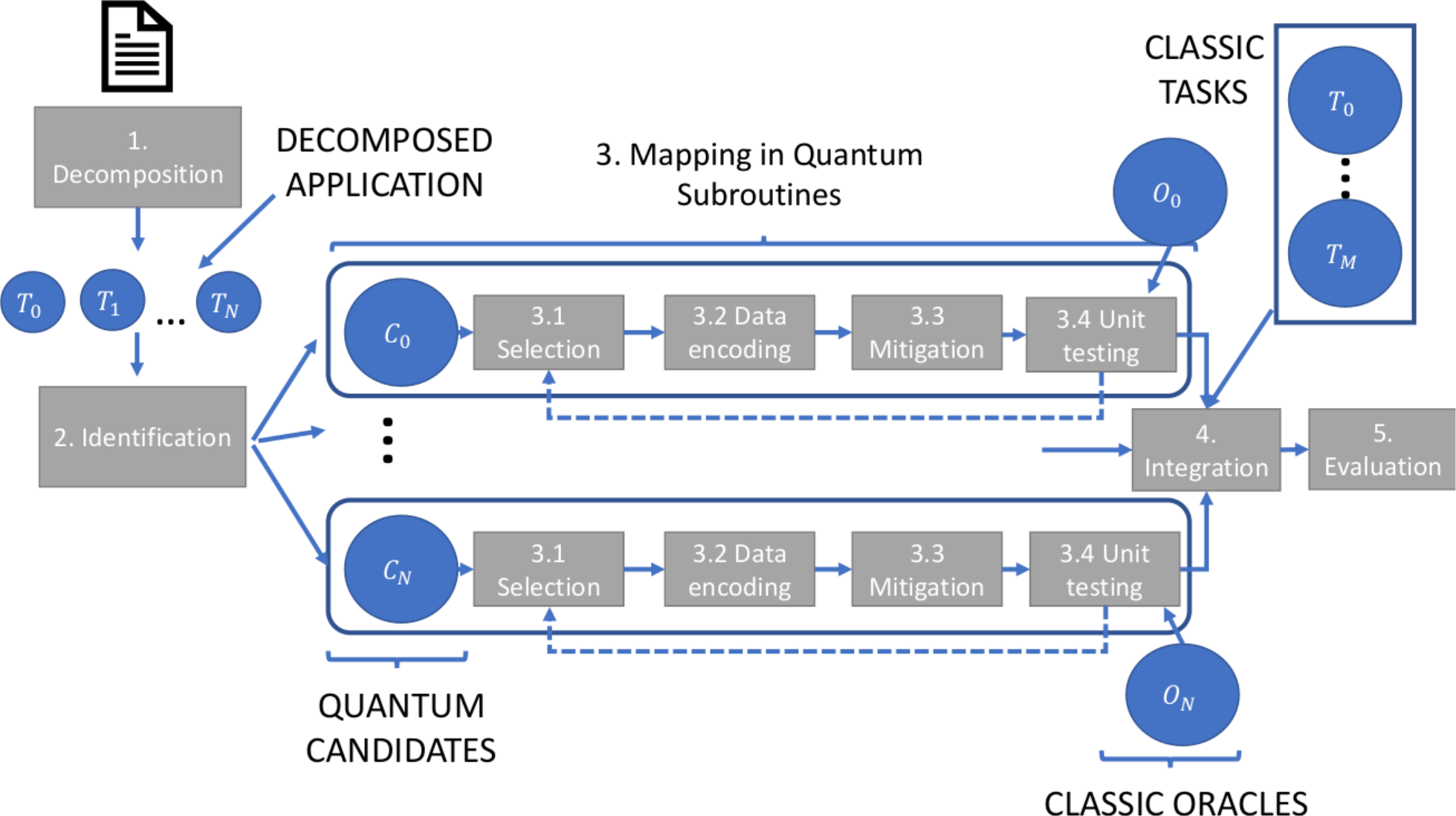}
\caption{Overview of a software evolution process.}
\label{fig:overview}
\end{figure}

 Figure~\ref{fig:overview} describes our software development process. First of all, the application is decomposed (conceptually) into different interdependent tasks. As a consequence, we describe our process starting from the identification of quantum candidates. The goal of this process is to identify the quantum candidates, i.e., tasks that can be executed on a quantum machine. The output is then a set $C$ of $N$ quantum candidates and a set $T$ of classic tasks. In step 2, each quantum candidate is mapped into a quantum subroutine, i.e., a quantum algorithm that is \emph{semantically equivalent}~\cite{klijnsma2023semantic,pitts2002operational} to the quantum candidate. Mapping is performed in four sub-steps: first, there is a \emph{selection} step, where a semantically equivalent algorithm is selected. If more algorithms are available for the same problem, selection is performed based on their resource requirements, available quantum machines or based on expert knowledge. Based on the selected algorithm, an appropriate data encoding method is applied, in order to translate data from the classic to the quantum domain, and an appropriate error mitigation method is selected, based on the performance of available methods on target quantum machines. Selection of data encoding, algorithm settings, and error mitigation methods can be performed by means of hyperparameter optimization methods, as in~\cite{Cranganore2022}. Finally, unit testing of the resulting quantum subroutine (inclusive of the encoding, the algorithm execution, and the error mitigation) is performed using the results of the classic implementation of the candidate as an oracle $O$. In case the error surpasses a user-defined threshold, the mapping process is restarted from the selection. In this work, we use Mean Square Error (MSE) as an error metric and select the configuration that minimizes MSE. Afterward, classic tasks and quantum subroutines are integrated by means of different software layers, and evaluation is performed using the complete classic application as an oracle.  

\subsection{Application Decomposition}
\begin{figure*}
\begin{subfigure}{0.49\textwidth}
    \centering
     \includegraphics[width=\columnwidth]{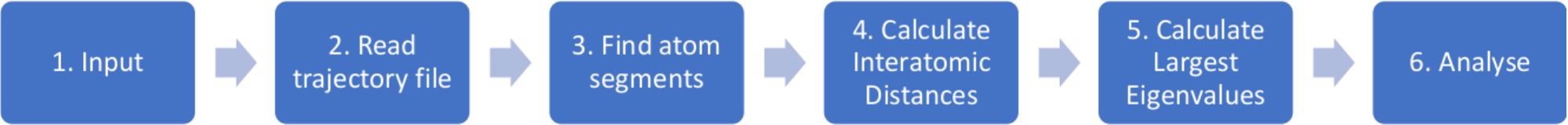}
     \caption{Classical MD Workflow~\cite{CRANGANORE2024346}.}
     \label{fig:workflow-classic}
     \end{subfigure}
     \begin{subfigure}{0.49\textwidth}
    \centering
     \includegraphics[width=\columnwidth]{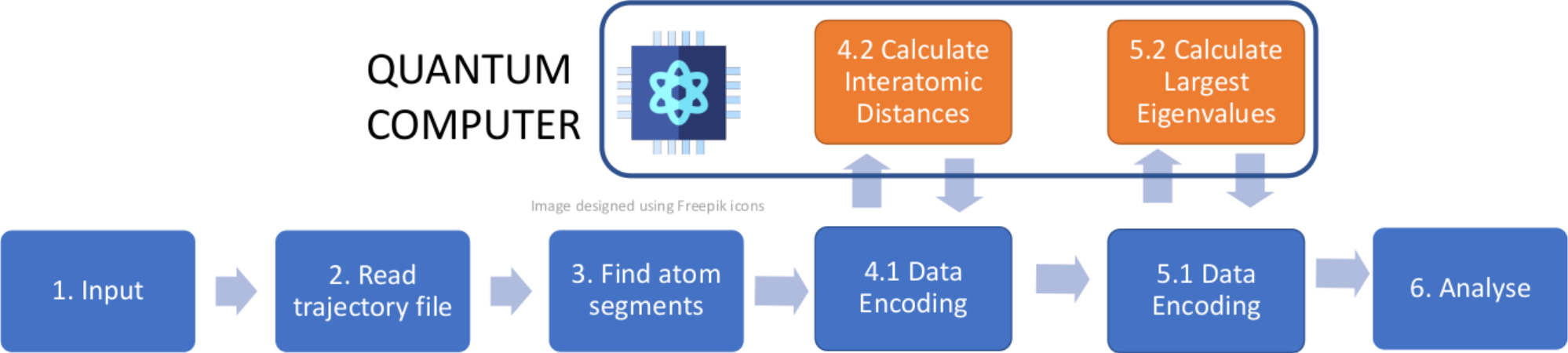}
     \caption{Hybrid MD Workflow~\cite{CRANGANORE2024346}.}
     \label{fig:workflow-quantum-decomposition}
     \end{subfigure}
     \caption{From Classical to Hybrid Quantum-Classical MD Workflow.}
\end{figure*}
\textbf{Methodology.}
The first step is decomposing the original classic application conceptually into a set of discrete tasks. The input of this process is a scientific program written in a high-level imperative language, including functions and subroutines that model the scientific process. This code is inspected with the help of experts of Team A to identify tasks that characterize the scientific program. In the target use case, the input application is an MD simulation, provided by Team A. The application has been provided as a scientific workflow, therefore it comes in the form of a DAG where each node is a task and each edge defines a precedence relationship between each task.

\noindent\textbf{Use Case Specifics.}
The application is composed of the following different tasks: (\textit{1}) \emph{Input phase}, where the user inputs simulation parameters; (\text{2}) \emph{Read trajectory file}, where the input trajectory file is read to extract information about evolution of the molecular system; (\textit{3}) \emph{evolution of atom segments}, where the evolution of the system is simulated; (\textit{4}) \emph{computation of inter-atomic distance}, where the new inter-atomic distances between different atom segments are calculated and stored in \emph{distance matrices}, $D$, such that each cell $d_{i,j}$ contains the Euclidean distances between atom segment $i$ and $j$; (\textit{5}) \emph{computation of largest eigenvalues}, where for each distance matrix we compute the largest eigenvalue, which is one of the so-called \emph{collective variables} capturing the evolution of the system. Since the scientific workflow is designed to be executed on Pegasus WMS~\cite{pegasusref,tanaka2022automating}, we asked Team A to modify it to be executed outside the WMS, to allow us to focus more on the quantum parts. Application decomposition is summarized by Figure~\ref{fig:workflow-classic}.


\subsection{Identification of Quantum Candidates}
\textbf{Methodology.} Afterwards, tasks are examined to identify \emph{quantum candidates}, i.e., tasks of the scientific application for which it is possible to find a quantum algorithm, called \emph{quantum subroutine}. Typical tasks that are not amenable to quantum execution usually involve I/O (e.g., reading input). For instance, Task 1 involves reading user input and setting up the simulation. Since data processing is not intensive in this phase, and also input operations are included, which cannot be performed efficiently on quantum hardware, this task is executed locally, on classic hardware, as well as task 2.

\noindent\textbf{Use Case Specifics.} Identification of quantum tasks was performed with the help of experts in Team B. Task 3 involves the selection and update of different atom segments, which can be performed very efficiently by applying vectorization on classic hardware. Task 4 requires calculating the Euclidean distance between atom segments, which has a very efficient implementation on quantum hardware~\cite{Cranganore2022}, the same as Task 5, which involves the computation of eigenvalues. Based on this analysis, we decided to execute tasks 4 and 5 on quantum machines.



\subsection{Mapping Quantum Candidates to Quantum Subroutines}
\textbf{Methodology.} Once quantum candidates have been identified, the next step is to map them into quantum subroutines, i.e., high-level code that can be executed by quantum machines. This process has different sub-steps: \textbf{(1) Algorithm selection: } Identifying a semantically equivalent quantum algorithm for the same problem, which is done in the selection phase together with the data encoding method; \textbf{(2) Data encoding: } Data coming from the classic domain have to be encoded to be used by the quantum algorithms. Different methods are available in literature, such as \emph{basis encoding}, \emph{amplitude encoding}, \emph{angle encoding}, and others~\cite{weigold2021encoding}, and each method affects application performance in different ways;  \textbf{(3) Error mitigation: } State-of-the-art quantum hardware is in the NISQ (Noisy Intermediate Scale Quantum) era, meaning, current quantum machines offer a limited amount of qubits (up to $127$), and are subject to noise due to the influence of surrounding environment and subject to quantum decoherence~\cite{Preskill2018quantumcomputingin}. As a consequence, different error correction methods need to be applied to the output of the quantum execution. Different methods can be applied in this context, such as ML-based~\cite{domingo2023taking}, autoencoders~\cite{Locher2023quantumerror} and error correction codes~\cite{sivak2023real}; \textbf{(4) Unit testing: } Since the output of the quantum subroutines might differ from the output of the classical code, unit test must be performed to evaluate the amount of error 

\noindent\textbf{Use Case Specifics.} We describe the use-case specifics for each of the identified steps. This required extensive communications with Team B, which has expertise in quantum computing.

\subsubsection{Algorithm Selection}
We describe the selection of quantum subroutines for Task 4 and 5.

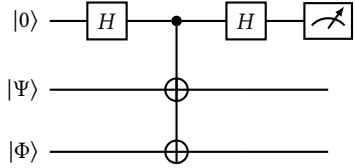
\begin{figure}[!ht]
    \centering
    \begin{quantikz}
    \lstick{\ket{0}} & \gate{H} & \ctrl{2} & \gate{H} & \meter{} \\
    \lstick{\ket{\Psi}} & \qw & \targ{} & \qw & \qw \\
    \lstick{\ket{\Phi}} & \qw & \targ{} & \qw & \qw & \\
    \end{quantikz}
    \caption{SWAP Test Circuit.}
    \label{fig:swap-test}
\end{figure}

SWAP test was identified as the best candidate for Euclidean distance between atom segments (Task 4). SWAP test is a quantum algorithm that expresses the inner product of two quantum states, $\ket{\psi}$, $\ket{\phi}$) in terms of the probability measurement of an auxiliary qubit, which acts as a controller qubit for the SWAP operator. The probability $P(\ket{0}) = 0.5$ of the auxiliary qubit means that the states are orthogonal, while the probability $P(\ket{0}) = 1$ means the states are identical. The circuit modelling the SWAP test is summarized by Figure~\ref{fig:swap-test}, while the Python code for SWAP Test is summarized by Figure~\ref{fig:swap-test-code}. In lines 1-5, we define three quantum registers, $qr_1$, $qr_2$ and $control_0$. The first two contain the encoding of the two atoms, while the control qubit is used to store the probability that two states $qr_1$ and $qr_2$ are orthogonal. Also, we initialize a classical register, $c$, where we store the result of measurements. These variables are used to initialize a \texttt{QuantumCircuit} object that models the quantum circuit performing the SWAP test. Afterward, $qr_1$ and $qr_2$ are initialized to the value of input atoms, $a$ and $b$, then, a Hadamard gate is applied to control qubit, to put it in a \textit{superposition}. Based on values of $control_0$, CSWAP gate is applied to $qr_1$ and $qr_2$ and a Hadamard gate is applied to $control_0$, that now contains the Euclidean distance between $a$ and $b$. Finally, $control_0$ is measured to obtain the Euclidean distance and stored in the classic register $c$, so that it can be used by the classic application.

Concerning the largest eigenvalue $lev$ calculation (Task 5), different algorithms could be used, such as HHL~\cite{nghiem2023quantum} and Variational Quantum Eigensolver~\cite{cerezo2022variational} (VQE).   

We selected VQE, which allows reducing the number of qubits needed to represent the input $m \times n$ matrix $BPM$ to $\log_2(m \cdot n)$. A very efficient VQE implementation is also already provided by Qiskit, which allowed to speed up the development. Source code is described in Figure~\ref{fig:vqe-code} in Appendix~\ref{sec:appendix}. 

VQE is defined by (1) the initial setting of $\Theta$ parameters, (2) a parametrized quantum circuit $C$ that is used for the preparation of input state $\psi(\Theta)$, (3) a cost function $\phi$ on parameters $\Theta$, and (4) an optimization algorithm. VQE cost function is defined as
\begin{equation}
    \Theta^* = \operatorname*{arg\,min}_\Theta C(\Theta) = \bra{\psi(\Theta)}H\ket{\psi(\Theta)},
    \label{eq:vqe:costfunction}
\end{equation}
meaning that for each state $\ket{\psi(\Theta)}$ one tries to minimize expectation of $H$. It follows from the variational principle that, $\operatorname*{arg\,min}_\Theta \phi(\Theta) \geq E_G$, therefore minimizing $ \phi(\Theta)$ enables us to approximate the ground state of $H$, $E_G$, where $\ket{\psi(\Theta)} = C(\Theta)\ket{\psi_i}$ corresponds to the minimum $H$ eigenvalues.

VQE takes as input three main parameters: the input matrix, $qubit\_op$ that corresponds to $H$; the $variational\_form$, that includes the circuit $C$ and the $\Theta$ parameters, and the $optimizer$, that is the optimization algorithms used by the framework to minimize the cost function in Equation~\ref{eq:vqe:costfunction}. Since we are interested in the maximum eigenvalue, we compute $-\mathtt{qubit\_op}$. Concerning the circuit $C$ and the optimization algorithm, they are selected from the Qiskit API, based on a parameter studio described in~\cite{Cranganore2022}.





\subsubsection{Data Encoding}
In some cases, such as VQE, the encoding is straightforward, since the algorithm requires specific data formats that allow to generalize encoding through specific classes and interfaces, i.e., in this case, \texttt{MatrixOp} class provided by Qiskit framework. However, for other cases, such as the SWAP test, it is left to the developer to identify the most suitable way to encode classic data into a quantum state. In our case, we have to encode the coordinates of the atoms in the circuit to calculate the Euclidean distance between them. The states $\ket{\phi}$ and $\ket{\psi}$ contain the coordinates of the atoms, which we encode by means of \emph{amplitude encoding}. In Figure~\ref{fig:amplitude-encoding}, we show how to encode data using amplitude encoding. 

\begin{figure}[H]
    \centering
    \footnotesize
    \begin{lstlisting}[language=Python]
    def phi_reg(self,A,B):
        phi = [np.array([self.norm(A,B)[0][i] / 
                    (self.norm_factor(A,B)[i]), - 
                    self.norm(A,B)[1][i] /
                    (self.norm_factor(A,B)[i])])
               for i in range(len(self.norm(A,B)[0]))]
        return phi

    def psi_reg(self,A,B):
        psi = []
        vec = self.vector_padding(A,B)
        for i in range(len(vec[0])):
            psi.append((vec[0][i]) / 
                (self.norm(A,B)[0][i] * np.sqrt(2)))
            psi.append((vec[1][i]) /
                (self.norm(A,B)[1][i] * np.sqrt(2)))
        
        return np.array(psi).
                reshape(len(A), 2 ** (vec[2]))
    \end{lstlisting}
    \caption{Amplitude Encoding (commit 029af9e)}
    \label{fig:amplitude-encoding}
\end{figure}

\begin{figure}[!h]
    \centering
    \footnotesize
    \begin{lstlisting}[language=Python]
    vqe_inputs = {
        'ansatz': ansatz,
        'operator': qubit_op,
        'optimizer': optimizer,
        'initial_point': 
                np.random.random(ansatz.num_parameters),
        'measurement_error_mitigation': True,
        'shots': 1024
        }

        job = provider.runtime.run(program_id='vqe',
                                     inputs=vqe_inputs,
                                     options=options)
     
    \end{lstlisting}
    \caption{Error Mitigation (commit 6094be7)}
    \label{fig:error-mitigation}
\end{figure}

\begin{figure}[!h]
    \centering
    \footnotesize
    \begin{lstlisting}[language=Python]
    def calc_eigval_quantum(bpm, ansatz, backend,
    optimizer):
        qubit_op = MatrixOp(-bpm)

        vqe_inputs = {
            'ansatz': ansatz,
            'operator': qubit_op,
            'optimizer': optimizer,
            'initial_point': 
                np.random.random(ansatz.
                num_parameters),
            'measurement_error_mitigation': True,
            'shots': 1024
        }
        options = {
            'backend_name': backend,
        }
        provider = IBMQ.get_provider('XXX', 
                'XXXX', 'main')
        job = provider.runtime
                .run(program_id='vqe',
                    inputs=vqe_inputs,
                    options=options)

        while job.status() != JobStatus.RUNNING 
            and job.status() != JobStatus.ERROR:
        pass
        if job.status() == JobStatus.ERROR:
            print(job.error_message())
            print(job.logs())

        start = time.time()
        res = job.result()
        end = time.time()
        return [-np.real(res['eigenvalue']),
                end - start]

    def calc_eigval_classic(bpm):
        levs = lin_alg.eigvalsh(bpm)[-1]
        return levs
    
    \end{lstlisting}
    \caption{Quantum-Classical Eigenvalues (commit 6094be7)}
    \label{fig:integration-eigenvalues}
\end{figure}

\begin{figure}[!h]
    \centering
    \footnotesize
    \begin{lstlisting}[language=Python]
    def extract_bpm(segs, dist_function):
        seg1 = segs[0][0]
        seg2 = segs[1][0]

        seg1_l = len(seg1)
        seg2_l = len(seg2)
        d = dist_function(seg1, seg2)
    
        dt = np.transpose(d)
        z1 = np.zeros((seg1_l, seg1_l))
        z2 = np.zeros((seg2_l, seg2_l))
        bpm = np.block([[z1, d], [dt, z2]])
        return bpm
    \end{lstlisting}
    \caption{Integration of BPM calculation (commit 6094be7)}
    \label{fig:integration-bpm}
\end{figure}

\subsubsection{Error Mitigation}
In the target application, error correction has been set to the default methods applied by Qiskit, setting the \texttt{measurement\_error\_mitigation} parameter to \texttt{True}, as shown in Figure~\ref{fig:error-mitigation}, since it allows minimizing MSE in comparison with each classic oracle.


\subsubsection{Unit Testing}
First, we randomly generate our atom segments, using Python random module.
This randomized testing procedure could also be extended through quantum fuzzing~\cite{quanfuzz}.
Afterward, we calculate the $D$ matrix using randomly generated segments as input for the classic and quantum versions. The classic task is used as a classic oracle, and we calculate the MSE. The same is done for the calculation of largest eigenvalues, where we also evaluate different hyperparameters for eigenvalue calculation and select the configuration that minimizes MSE. 

\subsection{Integration}
\textbf{Methodology.} Another important step is to integrate the quantum code in the application flow, allowing the application to switch between the two versions if quantum hardware is available. To this end, first, we abstract the code for each quantum task into a function, defining a classic and a quantum version. 
Figure~\ref{fig:integration-euclid} shows an example for calculating Euclidean distance. 

\begin{figure}[!h]
    \centering
    \footnotesize
    \begin{lstlisting}[language=Python]
    def calculate_distance_quantum(A, B, backend):
        cswap_circuit = CSWAPCircuit
                (1, 1, 3, 1, backend, 8192)
        quantum_ED = cswap_circuit.
                    execute_swap_test(A, B, noise_model)
        arr = np.array(quantum_ED)
        return arr

    def classic_euclidean_distance(A,B):
        arr = np.array(distance.cdist(A, B, 'sqeuclidean'))
        print(arr.shape)
        return arr
    \end{lstlisting}
    \caption{Quantum-Classical Euclidean Distance (commit 6094be7)}
    \label{fig:integration-euclid}
\end{figure}

\noindent\textbf{Use Case Specifics.} This is done respectively for Euclidean distance (Figure~\ref{fig:integration-euclid}) and for calculation of eigenvalues (Figure~\ref{fig:integration-eigenvalues}, Appendix).  
We can see that both tasks impact on the distance matrix $D$ (variable \texttt{bpm} in the code): since Euclidean distance is used to calculate each $D_{ij}$ cell, which contains Euclidean distance between atom segment $i$ and $j$; Also, eigenvalues are calculated using $D$ as input. To allow selection of the distance function, we further abstract the initialization of the $D$ into the method \texttt{extract\_bpm}, which takes as input the atom segments $A$ and $B$ and the distance function (classic or quantum) that should be used, as shown in Figure~\ref{fig:integration-bpm}. 
\begin{figure}[!h]
    \centering
    \footnotesize
    \begin{lstlisting}[language=Python]
    def apply_swap_test(a,b)
        q1 = QuantumRegister(self._aux, name='control_0') 
        q2 = QuantumRegister(self._qr1n, name='qr_1')
        q3 = QuantumRegister(self._qr2n, name='qr_2')
        c = ClassicalRegister(self._crn, name='c')
        qc = QuantumCircuit(q1,q2,q3,c)

        qc.initialize(a, q2[0:self._qr1n])
        qc.initialize(b, q3[0:self._qr2n])
        qc.barrier()
        qc.h(q1[0])
        qc.cswap(q1[0], q2[0], q3[0])
        qc.h(q1[0])
        qc.barrier()
        qc.measure(q1, c)

        return qc
    \end{lstlisting}
    \caption{SWAP Test Circuit Code (commit 029af9e)}
    \label{fig:swap-test-code}
\end{figure}

\begin{figure}[!h]
    \centering
    \footnotesize
    \begin{lstlisting}[language=Python]
    vqe = VQE(-qubit_op, variational_form, optimizer)
    ret = vqe.run(quantum_instance)
    vqe_result = np.real(ret['eigenvalue'])
    \end{lstlisting}
    \caption{LEV Calculation Source Code(commit 887a34b)}
    \label{fig:vqe-code}
\end{figure}
\subsection{Evaluating Correctness}


Once each quantum subroutine has been correctly evaluated by means of unit testing, and there are several ways to organize the variation between the original code and the quantum variant.
In this particular case study, we chose to introduce configuration options that activate either variant at the start of the program.
This allowed us to easily experiment with both versions, but also write test programs that only differ in how we call our application. 
From a software design perspective, we could envision more sophisticated abstractions (e.g., by employing a strategy pattern~\cite{strategypattern}).


\section{Implications for Quantum Software Engineering}
\label{sec:implications}

\begin{table*}[!th]
\scriptsize
    \centering
    \begin{tabular}{|C{3cm}|C{3cm}|C{3.5cm}|C{3.5cm}|}
    \hline
     \textbf{Steps} & \textbf{Papers} & \textbf{Summary} & \textbf{Implications} \\
     \hline
     \hline
        \multirow{3}{3cm}{\textbf{Candidate Identification}} &  \multirow{3}{3cm}{\cite{Cranganore2022,orus2019quantum,Matic2022,weder:22}} & \multirow{3}{3.5cm}{Manual identification based on expert knowledge.} & \multirow{3}{3.5cm}{\textbf{Semi-automatic Candidate Identification}}\\
        & & & \\
        & & & \\
        \hline
        \multirow{3}{3cm}{\textbf{Mapping Quantum Candidates}} & \multirow{3}{3cm}{\cite{Cranganore2022,orus2019quantum,Matic2022,vietz:22}} & \multirow{3}{3.5cm}{Specifically designed code for each identified quantum tasks.} & \multirow{3}{3.5cm}{\textbf{Transformation to Efficient Quantum Algorithms}}\\
         & & & \\
         & & & \\
        \hline
        \multirow{2}{3cm}{\textbf{Data Encoding}} & \multirow{2}{3cm}{\cite{weigold2021encoding,Cranganore2022,orus2019quantum,Matic2022}} & \multirow{2}{3.5cm}{Problem-specific data encoding} & \multirow{4}{3.5cm}{\textbf{Quantum Framework API Evolution}}\\
        & & & \\
        \cline{1-3}
        \multirow{2}{3cm}{\textbf{Error Mitigation}} & \multirow{2}{3cm}{\cite{domingo2023taking,Locher2023quantumerror,sivak2023real}} & \multirow{2}{3.5cm}{Data-driven methods} & \\ 
        & & & \\
        \hline
        \multirow{3}{3cm}{\textbf{Integration}} & \multirow{3}{3cm}{\cite{Cranganore2022,paltenghi:22,vietz:22,weder:22}} & \multirow{3}{3.5cm}{Application- and framework-specific bindings} & \multirow{3}{3.5cm}{\textbf{Managing Variants in Hybrid Quantum Programs}}\\
        & & & \\
        & & & \\
        \hline
        \multirow{2}{3cm}{\textbf{Evaluation of Correctness}} & \multirow{2}{3cm}{\cite{Cranganore2022,fortunato:22,garcia:21,ali:21,luo:22,Matic2022,paltenghi:22,pontolillo:22,wang:21,zhao:21}} & \multirow{2}{3.5cm}{Ad-hoc difference with classic output} & \multirow{2}{3.5cm}{\textbf{Automated Differential Testing}}\\
         & & & \\
        \hline
    \end{tabular}
    \caption{Comparison with the State of the Art.}
    \label{tab:implications}
\end{table*}

In this case study, we identified the sequence of steps required to obtain a hybrid quantum-classical program from a classical. Now, we identify how each step is covered in the state of the art, with the goal to identify future research directions for QSE. Our analysis is summarized by Table~\ref{tab:implications}. 

Regarding the \textbf{identification of quantum candidates}, it is performed by application developers or domain experts, who have the right knowledge both of the application domain and of quantum computing to identify which tasks can be executed on quantum computers to achieve a speedup. As a consequence, providing methods to analyze programs and support users in the identification of quantum candidates would significantly speed up the development of hybrid applications.

A similar discussion could be done for the \textbf{mapping of candidates in the quantum subroutines}, since identifying the most efficient quantum subroutine for a specific task also requires knowledge of quantum computing. As a consequence, methods to support users in the transformation of classic tasks into quantum subroutines should be provided, to support users in the development of efficient hybrid applications.

Different \textbf{data encoding} methods can be applied, depending on the target application and the number of qubits available. Considering the fast evolution of both quantum hardware, quantum APIs should be able to keep up with the new features provided by quantum hardware, providing encoding methods that can exploit the capabilities of underlying quantum architecture.

Regarding \textbf{error mitigation}, most of existing methods rely on AI models of quantum hardware noise, which are then used as input of specific error mitigation methods. Design of such models require collection of a huge amount of data by purposely-designed benchmarks, that needs to be constantly updated when a new quantum architecture is released. Quantum APIs should then provide methods to collect noise data to update noise models, allowing to easily integrate new quantum hardware in target systems.

\textbf{Integration} of classic and quantum hardware is also a big issue, since current hybrid applications are tightly coupled to the target quantum framework that is used for their development. To easily integrate a wider spectrum of quantum hardware, a mechanism to manage different quantum subroutines' implementations (i.e., for different architectures or frameworks) is required. 

Finally, \textbf{evaluation of correctness} of hybrid application is generally performed by comparing the quantum output with the classic output by means of different error metrics. As a consequence, testing of hybrid applications requires running both applications and collect intermediate results at different stage of execution and for each variant, which is a tedious task and is left to the application developer. Therefore, formal methods for automated differential testing are needed to ensure quality of hybrid applications.
While automated differential testing has seen extensive work in testing compilers~\cite{compilers1, compilers2, compilers3}, program analyzers and solvers~\cite{solvers1, solvers2}, and recently database engines~\cite{databases1}. We see our study as a catalyst to develop differential testing techniques for hybrid quantum applications. This brings additional challenges due to the stochastic nature of quantum computing. 



\subsection{Automated Reasoning Techniques for Quantum Candidate Identification}
\label{sec:semi-automatic}

Identifying quantum candidates in classical scientific programs requires domain knowledge from both computer scientists and quantum computing experts. As the transition from classical to quantum computing gathers momentum, it becomes increasingly vital to develop methodologies that facilitate the detection of potential quantum algorithms embedded within existing codebases. One plausible avenue for achieving this objective is to establish a correspondence mapping between classical algorithms with known quantum counterparts. By identifying such correspondences, we can potentially unravel quantum candidates hidden within the code. This mapping serves as a crucial bridge, allowing us to draw parallels between classical and quantum algorithmic structures. To effectively identify quantum candidates within the original source code, it is imperative to operate at a level of abstraction that captures the high-level algorithmic structure. This abstract representation facilitates reasoning about the code's underlying logic, thereby enabling the identification of the original algorithm. 


\subsection{Quantum Framework API Evolution}
\label{sec:api}
Frameworks like Qiskit are constantly evolving to harness the latest advancements in the field. These frameworks offer essential bindings for quantum execution, enabling developers to access cutting-edge quantum algorithms and hardware. However, this rapid evolution poses a challenge for developers who need to keep their code up-to-date with the latest API changes. Even during the six-month course of our case, we needed to adapt our code base to fit the evolving Qiskit interfaces. There are multiple streams of work that has been exploring ways to automatically update existing code to newer versions of (framework or library) APIs under the umbrella term API evolution~\cite{apievolution}. The question at the forefront of these efforts is whether it is possible to automate refactoring, allowing developers to seamlessly adapt to the fast-paced evolution of the Qiskit library. This endeavor holds the promise of streamlining quantum development and ensuring that quantum applications remain at the forefront of scientific and technological progress.

\subsection{Managing Variants}
\label{sec:variants}
One of our key findings underscores the necessity of retaining the original implementation from the classical program to facilitate automated evaluation through differential testing. Differential testing is vital for validating the correctness and performance of the quantum translation. However, to enable such testing, we must preserve parts of the classical codebase, even for quantum candidate tasks, presenting a challenge on how to seamlessly integrate and manage these two program variants.
During our case study, we addressed the challenge of managing variants by introducing a configuration flag. While this approach can be effective in distinguishing between program variants, it also introduces the potential for inconsistent states and complexity in the codebase. Given the limitations of configuration flags and the inherent complexities they may introduce, there is a pressing need for further research into improved software design principles for hybrid quantum systems. Such principles should enable the concurrent description of both classical and quantum variants, facilitating integration and maintenance. 


\subsection{Threats to Validity}

\textbf{Generalizability.} Case study research is highly contextual and personal in nature. Findings from this study may be challenging to generalize beyond the specific case under investigation. We have taken measures to provide a detailed description of the research context and have aimed to extract broader insights and implications where possible. Further, as part of our analysis, we identified and presented a more general software engineering methodology and present this separately from use case specific aspects in our findings. 

\noindent\textbf{Researcher Bias and Interpretive Subjectivity.} The presence of the first author as an active participant in the research context may have influenced the behaviors and actions of other participants. This subjectivity can introduce bias into the data collection and interpretation process. To mitigate this threat, the researcher maintained reflexivity by continually examining and acknowledging personal biases and their potential influence on data collection and analysis. Further, analysis involves interpretation and sense-making, which can be subjective. Different researchers might interpret the same data differently. To enhance the validity of interpretations, analysis was performed by multiple authors of the paper, and we have maintained transparency in our analytic process, including detailing the criteria and rationale for data interpretation.

\noindent\textbf{Documentation and Data Availability} The reliance on personal notes, software version control history, and computational notebooks for data collection may lead to incomplete or biased documentation. Some critical events or perspectives may not have been adequately recorded. We have attempted to address this limitation by cross-referencing multiple data sources and emphasizing the importance of comprehensive record-keeping.

In conclusion, while auto-ethnography as part of this case study offers a valuable perspective for understanding complex phenomena, it is essential to recognize and address these potential threats to the validity of our findings.

\section{Conclusion}
We conducted a six-month case study to investigate the process, evolution, and communication involved in the development of hybrid quantum applications. Starting from a MD use case, we identified an evolution process to transform the classical MD program into a hybrid quantum application. We report on our experience based on the triangulation of different data sources collected during the case, derive a more general software evolution process, and finally formulate implications for the software engineering research community.
We believe that this work would be useful in guiding the software engineering community in the development of software development patterns and tools supporting the development of hybrid quantum applications.

\begin{acks}
 This work has been partially funded through the Themis project (Trustworthy and Sustainable Code Offloading), Austrian Science Fund (FWF): Grant-DOI 10.55776/PAT1668223, the Standalone Project Transprecise Edge Computing (Triton), Austrian Science Fund (FWF): P 36870-N, and by the Flagship Project HPQC (High Performance Integrated Quantum Computing) \# 897481 Austrian Research Promotion Agency (FFG). We acknowledge the use of IBM Quantum Credits for this work. The views expressed are those of the authors, and do not reflect the official policy or position of IBM or the IBM Quantum team. 
\end{acks}

\section*{Reproducibility Statement}
The code and results for the experiments are publicly available in a GitHub repository\footnote{\url{https://github.com/vindem/quantumMD}}. Please note that the code you see in the main branch might differ from the snippets reported in the paper, due to compatibility issues with the newest versions of Qiskit.

\onecolumn
\begin{multicols}{2}
\bibliographystyle{ACM-Reference-Format}
\bibliography{quantum-se}
\end{multicols}
\end{document}